\documentstyle[ltwol,epsfig]{article}

\def\be{\begin{equation}}
\def\ee{\end{equation}}
\def\bea{\begin{eqnarray}}
\def\eea{\end{eqnarray}}

\begin{document}

\title{PENGUIN CONTRACTIONS AND FACTORIZATION IN $B \to K \pi$
  DECAYS}

\author{M. CIUCHINI}

\address{Dip. di Fisica, Univ. di Roma Tre 
and INFN, Sezione di Roma Tre,\\
Via della Vasca Navale 84, I-00146 Roma, Italy}   

\author{E. FRANCO, G. MARTINELLI, M. PIERINI AND
  L. SILVESTRINI$^*$}

\address{Dip. di Fisica, Univ. di Roma ``La Sapienza''  and INFN,\\
  Sezione di Roma, P.le A. Moro, I-00185 Rome, Italy.}  


\twocolumn[\maketitle\abstracts{We study $\Lambda_{QCD}/m_B$ corrections to
  factorization in $B \to K \pi$ decays. First, we analyze these decay
  channels within factorization, showing that, irrespectively of the
  value of $\gamma$, it is not possible to reproduce the experimental
  data. Then, we discuss $\Lambda_{QCD}/m_B$ corrections to these
  processes, and argue that there is a class of doubly Cabibbo
  enhanced non-factorizable contributions, usually called charming
  penguins, that cannot be neglected. Including these corrections, we
  obtain an excellent agreement with experimental data. Furthermore,
  contrary to what is obtained with factorization, we predict sizable
  rate asymmetries in $B^\pm \to K^\pm \pi^0$ and $B \to K^\pm
  \pi^\mp$.}]

\section{Introduction}
The theoretical understanding of non-leptonic two body $B$ decays is a
fundamental step for testing flavour physics and CP violation in the
Standard Model and for detecting signals of new
physics.~\cite{Ciuchini:2000de,Ciuchini:1997zp,barbieri} The
increasing accuracy of the experimental measurements at the $B$
factories~\cite{babar,belle} calls for a significant improvement of
the theoretical predictions.  In this respect, important progress has
been recently achieved by systematic studies of factorization made by
two independent groups.~\cite{li,beneke} These studies, while
confirming the physical idea ~\cite{previous} that factorization holds
for hadrons containing heavy quarks, $m_{Q} \gg \Lambda_{QCD}$, give
the explicit formulae necessary to compute quantitatively the relevant
amplitudes at the leading order in $\Lambda_{QCD}/m_{Q}$. At this
Workshop, many talks and posters~\cite{bcp4} have discussed in detail
the predictions for various nonleptonic $B$ decay channels using the
formalism of refs~\cite{li,beneke}. However, only perturbative
corrections to factorization can be computed using these two
approaches. The question which naturally arises is whether in practice
the power-suppressed corrections, for which quantitative estimates are
missing to date, may be phenomenologically important for $B$ decays.
This problem was previously addressed in
refs.~\cite{charming,pham,fleischer}.  In particular, the main
conclusion of refs.~\cite{charming} was that penguin contractions of
the leading operators of the effective weak Hamiltonian, $Q_{1}$ and
$Q_{2}$, although formally of ${\cal O}(\Lambda_{QCD}/m_{Q})$, may be
important in cases where the factorized amplitudes are either colour
or Cabibbo suppressed.  The most dramatic effect of these
non-factorizable penguin contractions manifested itself in the very
large enhancement of the $B \to K \pi$ branching ratios, as was also
emerging from the first measurements by the CLEO
Collaboration~\cite{cleo}.  In this case, the effect was caused by
Cabibbo-enhanced penguin contractions of the operators $Q^{c}_{1}$ and
$Q^{c}_{2}$, usually referred to as {\it charming penguins}. Since the
original publications, about three years ago, several other decay
channels have been measured~\cite{cleobr,babarbr,bellebr} and the
precision of the measurements is constantly improving in time.  With
respect to the previous analyses, it is now possible to attempt a more
quantitative study of charming penguin effects and of the corrections
expected to the factorized predictions.

In this talk\footnote{Contribution to BCP4, Ise-Shima, Japan,
  presented by L.S.}, we will focus on $B \to K \pi$
decays, where the effect of ${\cal O}(\Lambda_{QCD}/m_{Q})$
corrections to factorization is most striking. The interested reader
can find a more general analysis, including also $B \to \pi \pi$
decays, in ref.~\cite{noifuture}.

\section{Formalism}
The physical amplitudes for $B \to K \pi$ decays are more
conveniently written in terms of RG invariant parameters built using
the Wick contractions of the effective Hamiltonian~\cite{BS}.  In the
heavy quark limit, following the approach of ref.~\cite{beneke}, it is
possible to compute these RG invariant parameters using factorization.
The formalism has been developed so that it is possible to include
also the perturbative corrections to order $\alpha_{s}$.  An
alternative approach is provided by the formalism of ref.~\cite{li}.
The two methods differ in the treatment of the ${\cal O}(\alpha_{s})$
terms.  We present results obtained with the formalism of
ref.~\cite{beneke} only, with the addition of the
non-perturbative $\Lambda_{QCD}/m_{b}$ corrections to factorization.
A comparison with the formalism of ref.~\cite{li} will be presented
elsewhere. 

In the leading amplitudes, we have taken into account the SU(3)
breaking terms by using the appropriate decay constants, $f_K$ and
$f_\pi$ and form factors, $f_K(0)$ and $f_\pi(0)$.  As for
$\Lambda_{QCD}/m_{b}$ corrections, we have assumed instead $SU(3)$
symmetry and neglected Zweig-suppressed contributions.  In this
approximation, all the Cabibbo-enhanced $\Lambda_{QCD}/m_{b}$
corrections to $B \to K \pi$ decays can be reabsorbed in a single
parameter $\tilde P_{1}$.  This parameter includes not only the
charming penguin contribution, but also annihilation and penguin
contractions of penguin operators. It does not include leading
emission amplitudes of penguin operators ($Q_3$--$Q_6$) which have
been explicitly evaluated using factorization.  Had we included these
terms, this contribution would correspond to the parameter $P_1$ of
ref.~\cite{BS}. For simplicity, in the following we will continue to
refer to $\tilde P_{1}$ as the charming penguin contribution.

We proceed with the usual likelihood method, by extracting the input
quantities weighted by their distribution, which is assumed to be flat
for theoretical errors and Gaussian for experimental ones.  Averages
and standard deviations are then obtained by weighting the output
quantities by the likelihood factor \be {\cal L} = e^{- \frac{1}{2}
  \sum_{i} {(BR_{i} -BR_{i}^{exp})^{2}}/{ \sigma_{i}^{2}}} \, , \ee
where $\sigma_{i}$ are the standard deviations of the experimental
$BR$s $BR^{exp}_{i}$ given in table~\ref{tab:inputs}.  For more
details on the likelihood procedure, the reader is referred to
ref.~\cite{Ciuchini:2000de}, where all aspects are discussed at
length.

\begin{table}
\begin{center}
\caption{Input values used in the numerical analysis. The form factors
  are taken from refs.~\protect\cite{latticeff,qcdsrff}, the CKM parameters
  from ref.~\protect\cite{Ciuchini:2000de} and the BRs correspond to our
  average of CLEO, BaBar and Belle results
  \protect\cite{cleobr,babarbr,bellebr}. All the $BR$s are given in units of
  $10^{-6}$.}
\label{tab:inputs} 
\vspace{0.2cm}
\begin{tabular}{|c|c|}
\hline
$f_\pi(0)$& $0.27 \pm 0.08$ \\ \hline
$f_K(0)/f_\pi(0)$ & $1.2 \pm 0.1$ \\ \hline
$\rho $ & $0.224 \pm 0.038$ \\ \hline $\eta$ & $0.317 \pm 0.040 $ \\ \hline
$BR(B_d \to K^0 \pi^0)$ & $(10.4 \pm 2.6)$ \\ \hline  
$BR(B^+ \to K^+ \pi^0)$ & $(12.1 \pm 1.7)$ \\ \hline
$BR(B^+ \to K^0 \pi^+)$ & $(17.2 \pm 2.6)$ \\ \hline
$BR(B_d \to K^+ \pi^-) $& $(17.2 \pm 1.6)$ \\ \hline
\end{tabular}
\end{center}
\end{table}
\vspace*{3pt}

\section{Results}

\subsection*{Results with factorization}
We start by considering the case in which we stick to factorization
and take the CKM parameters $\vert V_{ub}\vert $ and $\gamma$ from
other experimental determinations.  Here and in all the other cases
where $\vert V_{ub}\vert $ and $\gamma$ are taken from the standard
Unitarity Triangle Analysis (UTA), we use as equivalent input
parameters the values of $\bar \rho$ and $\bar \eta$ given in
table~\ref{tab:inputs} from the analysis of
ref.~\cite{Ciuchini:2000de}. These values correspond to \be \gamma =
(54.8 \pm 6.2)^{0} \label{eq:gamma} \, . \ee To analyze $B \to K \pi$
decays, we only need $f_K(0)$.  Alternatively we may take only $\vert
V_{ub}\vert $ from the experiments and fit the value of $\gamma$.  In
the first case, the results are given in table~\ref{tab:one} labelled
as ``$\gamma$ UTA'' and show a generalized disagreement between
predictions and experimental data.  In the second case, the value of
$\gamma$ is fitted and the results are labelled as ``$\gamma$ free''.
In this case the disagreement is reduced for $BR(B^+ \to K^0 \pi^+)$
and $BR(B_d \to K^+ \pi^-) $, but it remains important for $BR(B_d \to
K^0 \pi^0)$ and $BR(B^+ \to K^+ \pi^0)$.  The pattern $BR(B^+ \to K^0
\pi^+)$:$BR(B_d \to K^+ \pi^-) $:$BR(B_d \to K^0 \pi^0)$:$BR(B^+ \to
K^+ \pi^0)$=2:2:1:1, which is suggested by the data, and is well
reproduced when the contribution of the charming penguins is large, as
discussed in the following, is lost in this case.  Moreover the fitted
value of $\gamma=(162\pm 13)^{0}$ is in striking disagreement with the
results of the UTA. Although one may question on the quoted
uncertainty of the UTA result, it is clearly impossible to reconcile
the two numbers.  Thus either there is new physics or
$\Lambda_{QCD}/m_{b}$ corrections are important. We now consider the
latter possibility.

 \begin{table} 
 \begin{center} 
 \caption{Values for $B \to K \pi$ $BR$s obtained in the approach of
   ref.~\protect\cite{beneke}. The results in the ``$\gamma$ UTA''
   column have been obtained using the values of $\bar \eta$ and $\bar
   \rho$ in Table \protect\ref{tab:inputs}. The results in the
   ``$\gamma$ free'' column have been obtained by fitting $\gamma$ as
   a free parameter. All the $BR$s are given in units of $10^{-6}$.}
 \label{tab:one} \vspace{0.2cm}
 \begin{tabular}{|c|c|c|}  \hline 
 $BR$ & $\gamma$ UTA & $\gamma$ free  \\ 
 \hline 
 $K^0 \pi^0$ & $5.9 \pm 0.2 $& $5.7 \pm 0.4 $ \\  
$K^+ \pi^0$ & $4.8 \pm 0.2 $&$ 9.1 \pm 0.5$\\ 
 $K^0 \pi^+$ & $11.7 \pm 0.5 $&$ 11.6 \pm 0.8 $ \\ 
$K^+ \pi^-$ & $9.8 \pm 0.4 $&$ 17.7 \pm 1.$\\ 
\hline  
 \end{tabular} 
 \end{center} 
\end{table} 
\vspace*{3pt}

\subsection*{Factorization with Charming penguins}
\begin{figure}
\center
\epsfig{figure=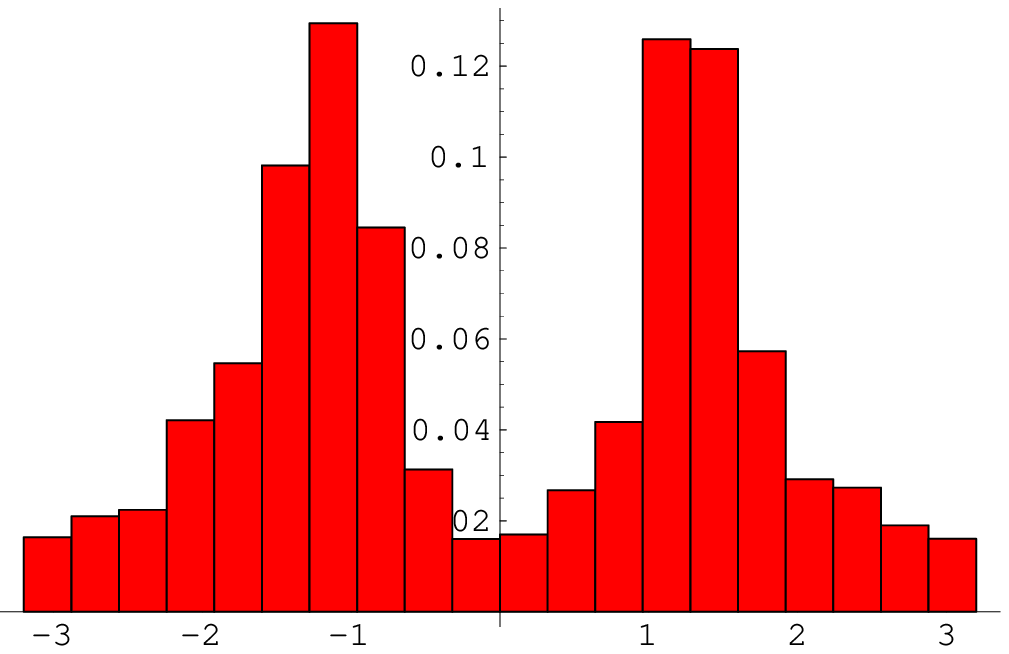,height=1.5in}
\caption{Probability density function for $\phi\equiv {\rm Arg} \tilde
  B_1$.}
\label{fig:phi}
\end{figure}

\begin{figure}
\center
\epsfig{figure=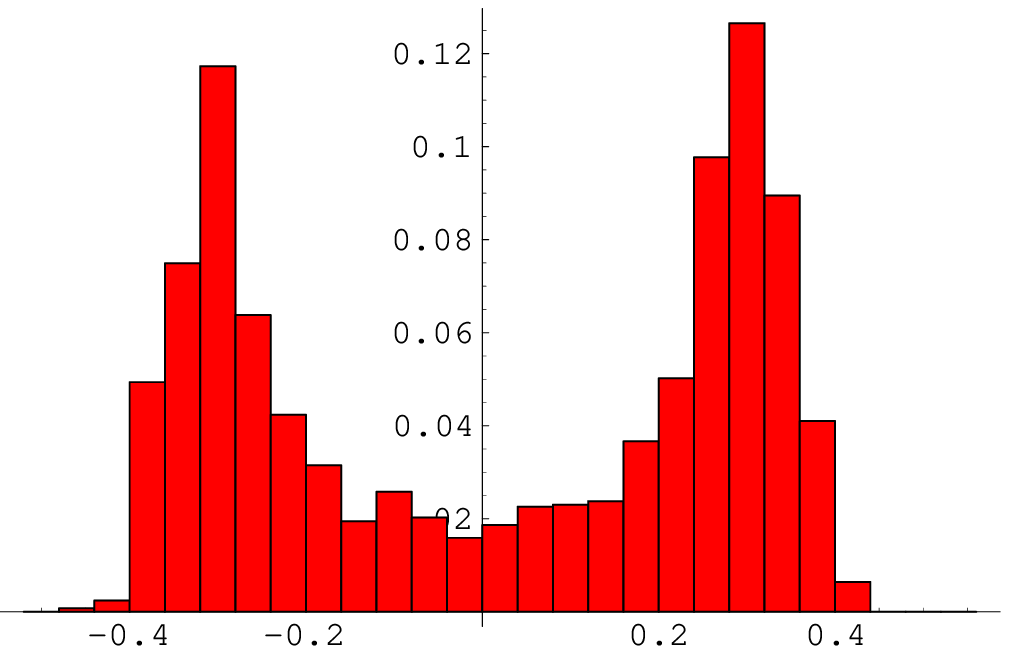,height=1.5in}
\caption{Probability density function for the rate CP asymmetry in $B^\pm
  \to K^\pm \pi^0$ decays.}
\label{fig:kpp0}
\end{figure}

\begin{figure}
\center
\epsfig{figure=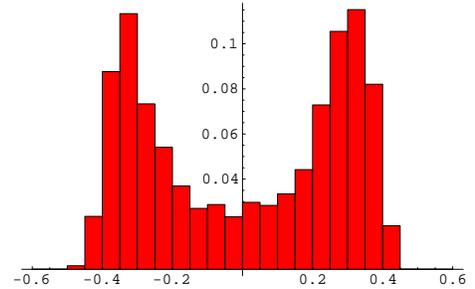,height=1.5in}
\caption{Probability density function for the rate CP asymmetry in $B
  \to K^\pm \pi^\mp$ decays.}
\label{fig:kppm}
\end{figure}

We now discuss the effect of the inclusion of charming penguins,
parametrized by $\tilde P_{1}$.  In general, this parameter is a
complex number and we fit it on the $B \to K \pi$ $BR$s.  In order to
have a reference scale for the size of charming penguins, we introduce
a suitable ``Bag'' parameter by writing \be \tilde P_{1} =
\frac{G_{F}}{\sqrt{2}} f_{\pi} f_{\pi}(0) g_{1} \tilde B_{1} \, , \ee
where $\tilde B_{1}$ is the $B$-parameter for $\tilde P_{1}$ and
$G_{F}$ the Fermi constant.  We always use $f_{\pi}(0)$ since, as
mentioned before, for these terms we work in the $SU(3)$ limit.
$g_{1}$ is a Clebsh-Gordan parameter which depends on the final $K
\pi$ channel.  In the case where $\vert V_{ub}\vert $ and $\gamma$ are
taken from the UTA, we find \be \vert \tilde B_{1} \vert = 0.12 \pm
0.04\, , \quad \quad \vert\phi\vert \equiv\vert{\rm Arg}(\tilde
B_{1})\vert = (81 \pm 37)^o \, . \ee There is a twofold ambiguity on
the sign of the phase which cannot be fixed by considering only
CP-averaged BRs, see fig.~\ref{fig:phi}. We will return on this point
in the following.  In table~\ref{tab:two} we give the corresponding
predicted values and uncertainties for the relevant branching ratios.
We observe a remarkable improvement in the fit.  Once charming
penguins are included, very little sensitivity to $\gamma$ is left and
therefore no information on $\gamma$ can be extracted from the study
of $B \to K \pi$ $BR$s.  However, the presence of a large phase in the
charming penguin contribution opens up the possibility of observing a
large rate CP asymmetry,
\begin{equation}
  \label{eq:a}
A=\frac{\Gamma(\bar B \to f) - \Gamma(B \to \bar f)}{\Gamma(\bar B \to
  f) + \Gamma (B \to \bar f)}\,,  
\end{equation}
 in $B \to K \pi$ decays. Indeed, while these
asymmetries come out to be always negligible if one uses the approach
of ref.~\cite{beneke}, when charming penguins are included we predict
visible asymmetries in two $B \to K \pi$ channels, as reported in
table~\ref{tab:two} and in figs.~\ref{fig:kpp0} and \ref{fig:kppm}.
Once the sign of the asymmetry is determined experimentally, the sign
ambiguity in the charming penguin phase can be resolved. With improved
experimental data, in the future one might think of extracting
informations on $\gamma$ from these asymmetries, exploiting the fact
that they are proportional to $\sin \gamma \sin \phi$ (neglecting the
very small perturbative strong phases).

 \begin{table} 
 \begin{center} 
 \caption{Values for $B \to K \pi$ $BR$s and rate CP asymmetries (in
   absolute value)
   obtained taking into account the contributions of charming
   penguins, and using the values of $\bar \rho$ and $\bar \eta$ in
   Table \protect\ref{tab:inputs}. All the $BR$s are given in units of
   $10^{-6}$.}
 \label{tab:two} 
\vspace{0.2cm}
 \begin{tabular}{|c|c|c|}  \hline 
 Channel & $BR$ & $\vert A \vert$  \\ 
 \hline 
 $K^0 \pi^0$ & $9.2 \pm 1.1 $& $0.0 \pm 0.0 $\\ 
$K^+ \pi^0$ & $11. \pm 1.6 $&$ 0.3 \pm 0.1$\\ 
 $K^0 \pi^+$ & $18.4 \pm 2.1 $&$ 0.0 \pm 0.0 $\\ 
$K^+ \pi^-$ & $17.5 \pm 1.3 $&$ 0.3 \pm 0.1$\\ 
\hline  
 \end{tabular} 
 \end{center} 
\end{table} 


\section{Conclusions}

We have studied $B \to K \pi$ decays, taking as a starting point the
factorized amplitude as obtained in the approach of
ref.~\cite{beneke}, and adding Cabibbo-enhanced nonperturbative
$\Lambda_{QCD}/m_B$ corrections (charming penguins), which turn out to
be dominant in these channels. We have shown that
\begin{itemize}
\item using factorization and the information on $\gamma$ from the
  UTA, all $B \to K \pi$ $BR$s come out to be much smaller than the
  experimental value;
\item even treating $\gamma$ as a free parameter, a sizable
  discrepancy between factorized predictions and experimental data
  remains present;
\item the inclusion of charming penguins, which is mandatory in these
  channels where they are doubly Cabibbo-enhanced, gives a perfect
  agreement with data irrespectively of the value of $\gamma$;
\item once charming penguins are included, no information on $\gamma$
  can be extracted from CP-averaged $B \to K \pi$ $BR$s;
\item sizable rate asymmetries are predicted for $B_d \to K^+ \pi^-$
  and $B^+ \to K^+ \pi^0$, and an experimental determination of the
  sign of the asymmetry can resolve the sign ambiguity on the charming
  penguin strong phase;
\item in the future, with more precise measurements, one might think
  of extracting informations on $\gamma$ 
  combining experimental values of CP-averaged $BR$s and rate
  asymmetries. 
\end{itemize}
For a more detailed discussion of these issues, together with a
careful analysis of $B \to \pi \pi$ decays, we refer the reader to
ref.~\cite{noifuture}. 

\section*{Acknowledgments}
We thank G. Buchalla, F. Ferroni and C.T. Sachrajda for discussions.
L.S. acknowledges very informative discussions with H.-n.  Li, and
thanks the organizers for the very lively and stimulating atmosphere
of the Workshop.

\end{document}